\RequirePackage[2018-12-01]{latexrelease}
\documentclass[showpacs,showkeys,11pt,
preprint,preprintnumbers,nofootinbib,
groupedaddress,superscriptaddress,amsmath,amssymb]{revtex4}
\usepackage{amsfonts}
\usepackage{amsmath}
\usepackage{graphicx,subfigure}
\usepackage{caption}
\usepackage[export]{adjustbox}
\usepackage{epsfig}
\usepackage{url}
\usepackage{multirow}
\usepackage{feynmp}

\newcommand {\be}{\begin{equation}}
\newcommand {\ee}{\end{equation}}
\newcommand {\ea}{\end{eqnarray}}

\begin{document}
\title{Observability of Parameter Space for Charged Higgs Boson in its bosonic decays in Two Higgs Doublet Model Type-1}

\author{Ijaz Ahmed}
\email{Ijaz.ahmed@fuuast.edu.pk}
\affiliation{Federal Urdu University of Arts, Science and Technology, Islamabad Pakistan}
\author{Waqas Ahmad}
\email{ahmadscientist77@gmail.com}
\affiliation{Riphah International University, Sector I-14, Hajj Complex, Islamabad Pakistan}
\author{ M. S. Amjad}
\email{sohailamjad@nutech.edu.pk}
\affiliation{National University of Technology, Islamabad Pakistan}
\author{Jamil Muhammad}
\email{mjamil@konkuk.ac.kr}
\affiliation{Sang-Ho College, and Department of Physics, Konkuk University, Seoul 05029, South Korea}

\begin{abstract}
This study explores the possibility of discovering $H^{\pm}$ through its bosonic decays, i.e. $H^{\pm}\rightarrow W^\pm\phi$ (where $\phi$ = h or A), within the Type-I Two Higgs Doublet Model (2HDM). The main objective is to demonstrate the available parameter space after applying the recent experimental and theoretical exclusion limits. We suggest that for $m_{H^\pm}$ = 150 GeV is the most probable mass for the $H^\pm\rightarrow W^\pm\phi$ decay channel in $pp$ collisions at $\sqrt{s}$ = 8, 13 and 14 TeV. Therefore we propose that this channel may be used as an alternative to $H^\pm\rightarrow \tau^\pm\nu$. 
 
\end{abstract}

\maketitle

\section{Introduction}
In 2HDM, $H^{\pm}$ is allowed to decay freely in fermions and gauge bosons. In Type-I, $ H^{\pm} \longrightarrow AW^{\pm}$ (decay to a neutral Higgs ``A" and ``W-boson") is a dominant channel.
The decay mode into $\tau^{+}\nu_\tau$ reaches branching ratios of more than 90$\%$ below the $t\bar{b}$ threshold and the muonic one ranges at a few $10^{-4}$ \cite{decaysupression1}. All other leptonic decay channels of the charged Higgs bosons are not important to be considered. As we know that there is a fermiophobic charged Higgs decay for large $tan\beta$ values, and it decays like $H^{\pm}\rightarrow W^{\pm}\phi$  ($ \phi=h, H, A$) if kinematically allowed and it would be dominant decay even for virtual $W^{\pm}$. Thus, it is the most encouraging mode of decay, for larger $tan\beta$ values. In Type-I 2HDM, the bosonic decays of light $H^{\pm}$ were recently studied at the LHC \cite{2017}. For decay processes $H^{\pm}\rightarrow W^{\pm}A$ and $H^{\pm}\rightarrow W^{\pm}h$, Branching Ratios are calculated in \cite{2017}. The $H^{\pm}\rightarrow W^{\pm}h$ reaches a BR of 10\% below the top-bottom threshold, at $tan\beta$ values from 2-3 and $m_{H^{\pm}}$ = 160 GeV.

For the production process, $pp \rightarrow tb H^{\pm}$, which is generally the most dominant mode for $H^{+}$, SM inclusive processes with top-quark pairs are inevitably an important background regardless of how $\phi$  decays \cite{alves2017charged}. Hence for more conventional 2HDM scenarios, the signal process from $\phi \rightarrow \tau\tau$ and $\phi \rightarrow bb$ provides promising avenue near the alignment limit \cite{coleppa2014charged,kling2015light}. Moreover in 2HDM, BR($\phi \rightarrow \tau\tau$) and BR($\phi \rightarrow bb$) with the relative size predicted under additional model assumptions, the search results from the two signatures may be combined to improve the sensitivity coverage of the 2HDM parameter space.
The {\tt 2HDMC-1.7.0} \cite{2hdmc} is used to put theoretical constraints and experimental bounds are applied. For that purpose, HiggsBounds\cite{higgsbounds} and HiggsSignals\cite{higgssignals} libraries are interfaced with 2HDMC, and also ScannerS\cite{scanners} is used to put the most recent experimental bound on the selected parameters and compare whether it is allowed or not experimentally.

\section{Review of 2HDM}
The scalar potential of 2HDM \cite{2HDMref} has 14 parameters, including the charge violation, and CP violation. The general term for scalar potential is as follows,
\newline
\begin{multline}
V=m^2_{11} \Phi^{\dagger}_1 \Phi_1 + m^2_{22} \Phi^{\dagger}_2 \Phi_2 - m^2_{12} \biggl\{ \Phi^{\dagger}_1 \Phi_2 + h.c. \biggr\} + \frac{\lambda_1}{2} (\Phi^{\dagger}_1 \Phi_1)^2 + \\
 \frac{\lambda_2}{2} (\Phi^{\dagger}_2 \Phi_2)^2 + \lambda_3 (\Phi^{\dagger}_1 \Phi_1)(\Phi^{\dagger}_2 \Phi_2) + \lambda_4 (\Phi^{\dagger}_1 \Phi_2) (\Phi^{\dagger}_2 \Phi_1) + \\
  \biggl\{ \frac{\lambda_5}{2} (\Phi^{\dagger}_1 \Phi_2)^2 + \left[ \lambda_6 (\Phi^{\dagger}_1 \Phi_1) + \lambda_7 (\Phi^{\dagger}_2 \Phi_2) \right] (\Phi^{\dagger}_1 \Phi_2) + h.c. \biggl\}
\end{multline}
where $( \lambda_{i}, i = 1,2,3,...., 7)$  are dimensionless coupling parameters, $ m^2_{11} , m^2_{22}$ and $m^2_{12}$ are squares of masses. To treat the 2HDM potential as charge and parity conserving potential, all the parameters should be real. The vacuum expectation value VEV is acquired by each scalar-doublet when electroweak symmetry breaks. The two doublets are,

\begin{equation}
<\Phi_1>=\binom{0}{\frac{v_1}{\sqrt{2}}} , <\Phi_2>=\binom{0}{\frac{v_2}{\sqrt{2}}}
\end{equation}
These two doublets lead to eight fields among which three correspond to massive $W^{\pm}$ and $Z^0$ vector bosons, and the remaining five fields lead to five physical Higgs bosons.
\begin{equation}
\Phi_i=\binom{\Phi^+_i}{\frac{(v_i+\rho_i+\iota\eta_i)}{\sqrt{2}}}
\end{equation}
where   i=1,2
with \begin{math} V_1=Vcos\beta, V_2=Vsin\beta \hspace{1mm} and \hspace{1mm} V_1,V_2 \geq 0 \end{math}
\newline
It satisfy the condition $V_{SM} = \sqrt{v^2_1+v^2_2}$. The experimentally obtained value of $V_{SM}$ is  246.22 GeV. The obtained fields are given as,
\begin{equation}
\binom{\rho_a}{\rho_b}= \begin{pmatrix}
cos\alpha & -sin\alpha\\sin\alpha & cos\alpha
\end{pmatrix} \binom{H}{h} \hspace{1mm},\hspace{1mm} \binom{\eta_a}{\eta_b} = \begin{pmatrix}
cos\beta & -sin\beta\\sin\beta & cos\beta
\end{pmatrix} \binom{G^0}{A}
\end{equation}
and
\begin{equation}
\binom{\Phi^{\dagger}_a}{\Phi^{\dagger}_b}= \begin{pmatrix}
cos\beta & -sin\beta\\sin\beta & cos\beta
\end{pmatrix} \binom{G^{\dagger}}{H^{\dagger}}
\end{equation}
Mass-matrix of charged Higgs states are diagonalized by rotational angle and it is defined as $tan\beta = \frac{V_{2}}{V_{1}}$. Similarly mass-matrix of scalar Higgs states are diagonalized by the rotational angle \begin{math} \alpha \end{math} and satisfies following relation,

\begin{equation}
\tan(2\alpha)=\frac{2(-m^2_{12}+\lambda_{345}V_1V_2)}{m^2_{12}(V_2/V_1 - V_1/V_2)+\lambda_1V^2_1-\lambda_2V^2_2}
\end{equation}
where,\begin{math} \hspace{1cm} \lambda_{345}=\lambda_3+\lambda_4+\lambda_5 \end{math}


\subsection{Theoretical and Experimental Bounds}
The general potential of 2HDM is too complex as compared to the one in the standard model SM. The 2HDM imposes theoretical constraints on the potential to guarantee the stability of potential.\newline
The Higgs potential should be positive throughout the field space. This ensures that a stable vacuum configuration for asymptotically large field values is maintained. The quartic terms are the leading terms at large field space values. The following substitutions are helpful,

\begin{equation}
\mid \Phi_1 \mid=r cos \phi \hspace{1mm},\hspace{1mm} \mid\Phi_2\mid=rsin\phi \hspace{4mm}and\hspace{4mm} \frac{\Phi_1\Phi^{\dagger}_2}{\mid\Phi_1\mid\mid\Phi_2\mid} =\rho\exp(\iota\theta)
\end{equation}
\newline
where, \hspace{1cm} \begin{math} \rho=\mid0-1\mid \hspace{1mm},\hspace{1mm} \theta=\mid0-2\pi\mid \hspace{3mm}and\hspace{3mm} \phi=\mid0-\frac{\pi}{2}\mid \end{math}

After making these substitutions, and omitting the common factor $r^4$, the quartic terms of the potential can be expressed as, 
\begin{multline}
V_{(4)}=\frac{1}{2}\lambda_1 cos^4\phi + \frac{1}{2}\lambda_2 sin^4\phi + \lambda_3cos^2\phi sin^2\phi+\lambda_4\rho^2cos^2\phi sin^2\phi \\
+ \lambda_5\rho^2 cos^2\phi sin^2\phi cos2\theta + [\lambda_6 cos^2\phi + \lambda_7 sin^2\phi]2\rho cos\phi sin\phi cos\theta
\end{multline}
From above equation we can see that the potential will be positive if,
\begin{equation}
\lambda_1>0 \hspace{4mm},\hspace{4mm}\lambda_2>0\hspace{4mm}and\hspace{4mm}\lambda_3>-\sqrt{\lambda_1\lambda_2}
\end{equation}
If \begin{math} \lambda_{6} = 0 = \lambda_{7} \end{math}, a case for natural conservation of flavor, then an extra condition must be satisfied,
\begin{equation}
\lambda_3+\lambda_4-\mid\lambda_5\mid>-\sqrt{\lambda_1\lambda_2}
\end{equation}
If either \begin{math} \lambda_{6} \not= 0 \hspace{1mm} or \hspace{1mm} \lambda_{7} \not= 0 \end{math} then the above condition changes to,
\begin{equation}
\lambda_3+\lambda_4-\mid\lambda_5\mid>-\sqrt{\lambda_1\lambda_2}
\end{equation}
\newline
Scattering matrices are unitary in order to conserve probability. In the theory of weak couplings, the contribution of higher order terms decreases gradually. While in the theory of strong couplings, individual contributions increase arbitrarily. The eigenvalues (\begin{math} L_i \end{math} ) of S-matrices must satisfy the condition \begin{math} L_i \leq 16\pi \end{math} in order to achieve the tree-level unitarity that means the saturation of S-matrices up to tree-level unitarity.\newline
Perturbation constraints requires that the quartic Higgs couplings must satisfy the condition \begin{math} \mid C_{H_{i},H_{j},H_{k},H_{l}} \mid \leq 4\pi \end{math}. One can imagine that some interaction channels are non-perturbative while others are perturbative. \begin{math} \mid \lambda_i \mid = 4\pi\xi \end{math} , is another way to explain this constraint, where \begin{math} \xi = 0.8 \end{math} \footnote{The value is arbitrarily chosen as an upper bound}. This gives $\mid \lambda_i \mid \leq 10$ for $\lambda_i$, as the upper bound. 

Alongside theoretical constraints, there are also experimental constraints coming from B-Physics and various experiments on different colliders from recent Higgs searches. Here we discuss some important constraints. Also, using SuperIso V.3.2\cite{mahmoudi2008superiso}, we list the SM prediction values provided in this category. The Standard Model BR for $(B_\mu \rightarrow \tau\nu)_{SM}$ reported in \cite{mahmoudi2008superiso} is: 

\begin{equation}
BR(B_\mu\rightarrow\tau\nu)_{SM} \hspace{4mm}=\hspace{4mm}(1.01\hspace{2mm} \pm \hspace{2mm} 0.29)\hspace{2mm} \times \hspace{2mm} 10^{-4}
\end{equation}
The standard model estimation may, in fact, be contrasted to the most recent heavy flavour averaging (HFAG) result.\cite{asner2010averages}. 
\begin{equation}
BR(B_\mu\rightarrow\tau\nu)_{exp} \hspace{4mm}=\hspace{4mm}(1.64\hspace{2mm} \pm \hspace{2mm} 0.34)\hspace{2mm} \times \hspace{2mm} 10^{-4}
\end{equation}
so the ratio will become,
\begin{equation}
R^{exp}_{SM}\hspace{4mm}=\hspace{4mm}\frac{BR(B_\mu\rightarrow\tau\nu)_{exp}}{BR(B_\mu\rightarrow\tau\nu)_{SM}} \hspace{4mm}=\hspace{4mm}1.62\hspace{2mm}\pm\hspace{2mm}0.54
\end{equation}
This causes the exclusion of two sectors of \begin{math} (tan\beta)/m_{H^\pm} \end{math} ratio in 2HDM \cite{mahmoudi2010flavor}. This implies that for \begin{math} tan\beta \geq 1 \end{math}, the mass of charged Higgs must be greater than 800 GeV for Type-II 2HDM\cite{800gev2hdm}. As we know that the ( \begin{math} B_\mu \rightarrow {\tau}{\nu}_{\tau} \end{math} ) decay depends on \begin{math} \mid V_{ub} \mid \end{math} so the  ( \begin{math} B_\mu \rightarrow Dl \nu \end{math} ) (semi-leptonic) decay also depends on \begin{math} \mid V_{ub} \mid \end{math} , i.e. more precisely known as compared to \begin{math} \mid V_{ub} \mid \end{math}  and the branching ratio of ( \begin{math} B_\mu \rightarrow {\tau}{\nu}_{\tau} \end{math} ) is fifty times greater than the branching ratio of (\begin{math} B_\mu \rightarrow {\tau}{\nu} \end{math}) in standard model but it is still difficult to detect because two neutrinos exist in its final state. The 2HDM deals only with the numerator of the ratio.

\begin{equation}
{\xi}_{Dl\nu_\tau} \hspace{4mm}=\hspace{4mm} \frac{BR(B\rightarrow D{\tau}{\nu}_{\tau})}{BR(B\rightarrow Dl{\nu}_{\tau})}
\end{equation}
and allow us to reduce to some extent the theoretical uncertainties. The experimental outcomes by BaBar collaborations and SM predictions \cite{mahmoudi2010flavor} are as follows.

\begin{equation}
{\xi}^{SM}_{Dl{\nu}_{\tau}} \hspace{4mm}=\hspace{4mm} (29.7 \hspace{2mm} \pm \hspace{2mm} 3)\hspace{2mm} \times \hspace{2mm} 10^{-2}
\end{equation}
\begin{equation}
{\xi}^{exp}_{Dl{\nu}_{\tau}} \hspace{4mm}=\hspace{4mm} (44.0 \hspace{2mm} \pm \hspace{2mm} 5.8 \hspace{2mm} \pm \hspace{2mm} 4.2)\hspace{2mm} \times \hspace{2mm} 10^{-2}
\end{equation}
\newline
For ($B \rightarrow X_s\gamma$), this special transition is mediated by \begin{math} H^{\pm} \end{math} and it includes Flavor-changing neutral current (FCNC) and \begin{math} W^{\pm} \end{math} contributions.
As to the respective BR, the contribution of charged Higgs is always positive so to probe Type-II 2HDM, this can be used efficiently. For this transition, the NNLO-SM predicted $(3.34\pm0.22)\times10^{-4}$ for $BR(B \rightarrow X_s\gamma)_{SM}$ \cite{mehmodi-2008}.
So, for $BR(B \rightarrow X_s\gamma)_{exp}$ recently experimentally calculated value is $(3.32\pm0.15)\times10^{-4}$,
For Type-II Yukawa interactions, this constraint excludes the light-charged Higgs. Higher order analysis \cite{Borzumati:1998nx} estimated lower limit of \begin{math} M_{H^{\pm}} \end{math} is 380 GeV at 95\% C.L. However, it is important to mention that the bound in \cite{Borzumati:1998nx} does not include novel experimental and theoretical predictions and hence the numerical results maybe outdated.\newline

For ($D_S \rightarrow \tau\nu$), the SM  prediction is $(3.32\pm0.15)\times10^{-4}$ \cite{mahmoudi2008superiso}, for \begin{math} f_{Ds} = 0.248 \pm 2.5 GeV \end{math} \cite{PhysRevD.82.114504} and the updated experimental calculation for $BR(D_s \rightarrow \tau\nu)_{exp}$ is $(5.51\pm0.24)\times10^{-2}$ \cite{theheavyflavoraveraginggroup2011averages}, for ($ B_{d/s} \rightarrow \mu^+\mu^- $). At large $tan \beta$ values, the lower limit on charged Higgs mass $m_{H^{\pm}}$ is given in \cite{logan2000bs}.
For decays $BR(B_s \rightarrow {\mu}^+{\mu}^+)_{SM}$ and $BR(B_d \rightarrow {\mu}^+{\mu}^+)_{SM}$, SM predictions are $(3.54\pm0.27)\times10^{-9}$ and $(1.1\pm0.1)\times10^{-9}$ respectively \cite{mahmoudi2008superiso}. Experimental results for the limits of these decays at 95\% C.L. are $BR(B_s \rightarrow {\mu}^+{\mu}^+)_{exp} < 4.5 \times 10^{-9}$ and $BR(B_d \rightarrow {\mu}^+{\mu}^+)_{exp} < 1.0 \times 10^{-9}$ presented by LHCb collaboration \cite{aaij2012strong}. If the results from ATLAS and CMS \cite{lhcbatlas} in above limits are also added then more strict limits are $BR(B_s \rightarrow {\mu}^+{\mu}^+)_{exp} < 4.2 \times 10^{-9}$ and $BR(B_d \rightarrow {\mu}^+{\mu}^+)_{exp} < 8.1 \times 10^{-10}$.

\section{Discussion}
\begin{figure}[!hbt]

{\includegraphics[width=7.5cm]{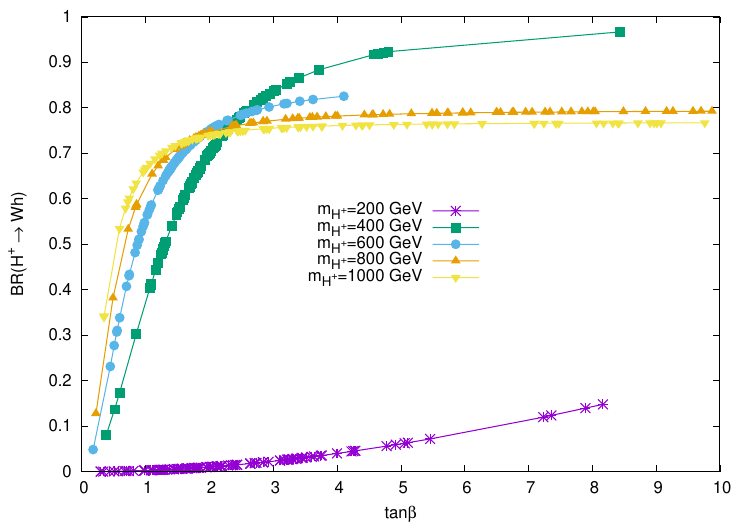}
\label{BP1Whtheo.pdf}
}\hfill
{\includegraphics[width=7.5cm]{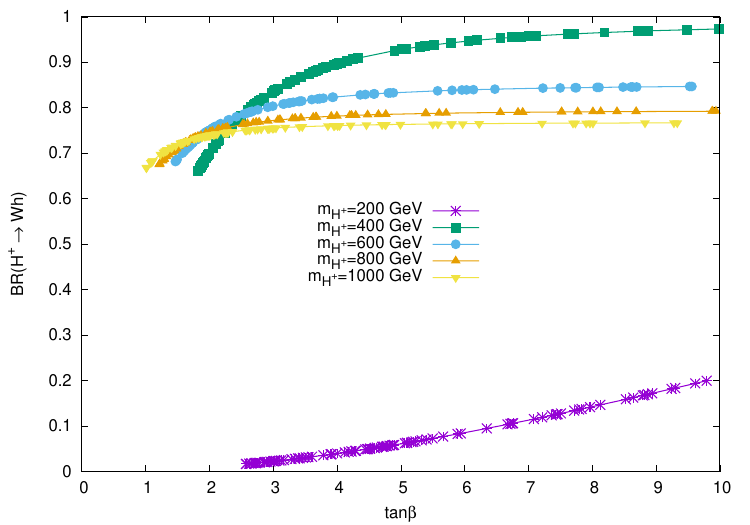}
\label{BP1Whexp.pdf}
}\hfill
\caption{In Type-I 2HDM, $tan\beta$ vs BR($H^{+}\rightarrow W^{+}h$) is scanned in the face of theoretical (left) and experimental (right) restrictions (constraints) with $m_{h}$=125 GeV, $m_{H}$=300 GeV, $m_{H^{+}}$=$m_{A}$=200-1000 GeV, $sin(\beta-\alpha)=$0.85, we set ${m^{2}}_{12}=\frac{[{m^2}_{H^+}][tan{\beta}]}{[1+{tan{\beta}}^2]}$.}
\label{BRHtoWPhi}
\end{figure}

The branching ratios are calculated using HDECAY \cite{hdecay} through the anyHdecay interface of ref \cite{scanners}, and predictions for gluon-fusion and bb-associated Higgs production at hadron colliders are obtained using tabulated results from SUSHI \cite{sushi}. The V H-associated (sub)channel cross section predictions are made using the HiggsBounds parametrizations, and the charged Higgs production in association with a top-quark is tested using the HiggsBounds  and HiggsSignals.
These above constraints and calculations are implemented in ScannerS.
Figure \ref{BRHtoWPhi}, shows BR\begin{math}(H^{+}\rightarrow W^{+}h) \end{math} with respect to $\tan \beta$ values for different masses of $m_{H^+}$. One can see that for $m_{H^{+}}=800$ and $1000$ \hspace{1mm} GeV, the BR\begin{math}(H^{+}\rightarrow W^{+}h) \end{math} remains between $ 70\% \to 80\%$. For $m_{H^+}=600 \, \textrm{GeV}$, the maximum BR reaches more than $80 \%$ for large $\tan \beta$ values. For $\tan \beta > 3$, the maximum BR is achieved for $m_H^+ = 400$ GeV. For $m_{H^+}=200 \, \textrm{GeV}$, the BR remains less than $20\%$ across the $\tan \beta$ range.

\begin{figure}[!hbt]

{\includegraphics[width=7.5cm]{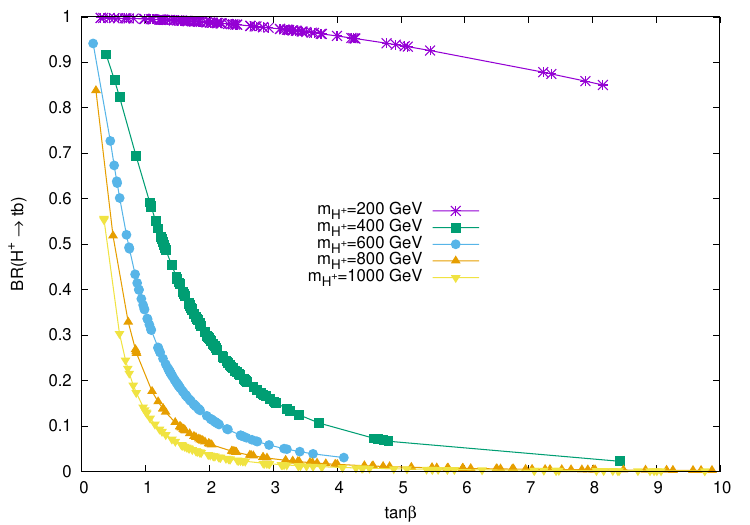}
\label{BP1tbtheo.pdf}
}\hfill
{\includegraphics[width=7.5cm]{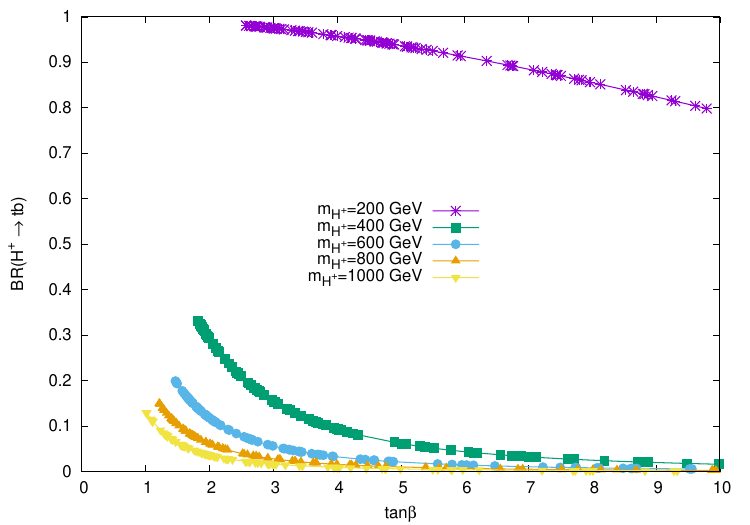}
\label{BP1tbexp.pdf}
}\hfill
\caption{In Type-I 2HDM, $tan\beta$ vs BR($H^{+}\rightarrow$ tb) is scanned in the face of theoretical (left) and experimental (right) restrictions (constraints) with $m_{h}$=125 GeV, $m_{H}$=300 GeV, $m_{H^{+}}$=$m_{A}$=[200-1000] GeV, $sin(\beta-\alpha)$=0.85, we set ${m^{2}}_{12}$=$\frac{[{m^2}_{H^+}][tan{\beta}]}{[1+{tan{\beta}}^2]}$.}
\label{BRHtoTB}
\end{figure}
Figure \ref{BRHtoTB} shows that, for \begin{math} m_{H^{+}}=200 \end{math}\hspace{1mm} GeV at \begin{math} 2\leq tan\beta \leq 3 \end{math}, the BR\begin{math}(H^{+}\rightarrow tb) \end{math} is the most dominant i.e. BR\begin{math}(H^{+}\rightarrow tb) \approx 100\% \end{math}. One can see from Figure \ref{BRHtoWPhi} and \ref{BRHtoTB} that for smaller $tan \beta$ values, the $H^+ \to tb$ decay mode is most dominant while for higher values, the $H^+ \to Wh$ becomes most probable. However, for $m_{H^+} = 200$ GeV, the scenario is different, as BR\begin{math}(H^{+}\rightarrow tb) \end{math}  is dominant while  BR\begin{math}(H^{+}\rightarrow Wh) \end{math} is less probable.


\begin{figure}[!hbt]

{\includegraphics[width=7.5cm]{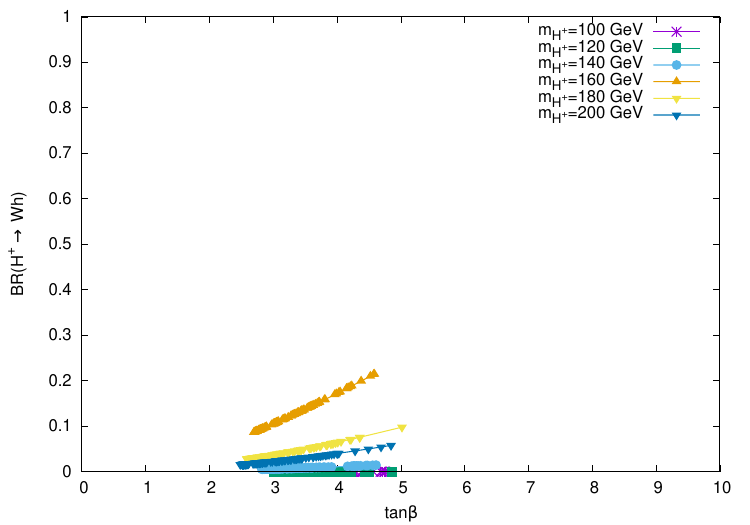}
\label{BP1Whexptheo.pdf}
}\hfill
{\includegraphics[width=7.5cm]{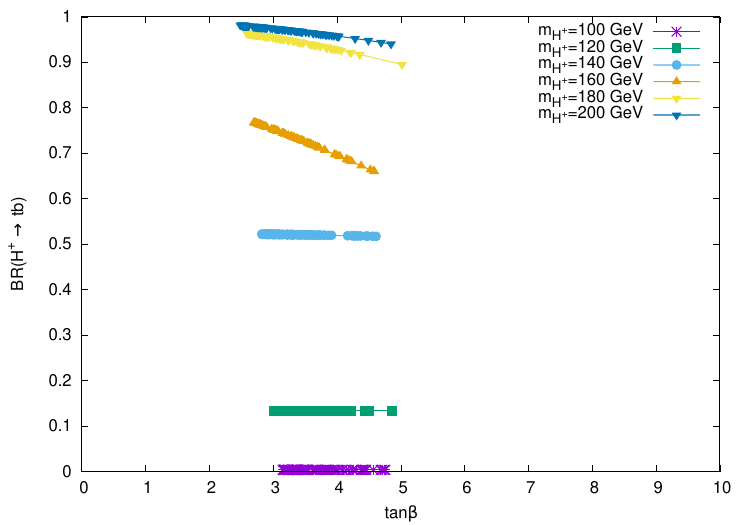}
\label{BP1tbexptheo.pdf}
}\hfill
\caption{In Type-I 2HDM, $tan\beta$ vs BR($H^{+}\rightarrow W^{+}h$) (left) $tan\beta$ vs BR($H^{+}\rightarrow t\bar{b}$) (right) is scanned in the presence of theoretical constraints as well as experimental constraints with $m_{h}$ = 125 GeV, $m_{H}$=300 GeV, $m_{H^{+}}$=$m_{A}$=[100-200] GeV, $sin(\beta-\alpha)$ = 0.85, we set ${m^{2}}_{12}=\frac{[{m^2}_{H^+}][tan{\beta}]}{[1+{tan{\beta}}^2]}$.}
\label{BRWhandtbConstraints}
\end{figure}

\begin{figure}[!hbt]
{\includegraphics[width=7.5cm]{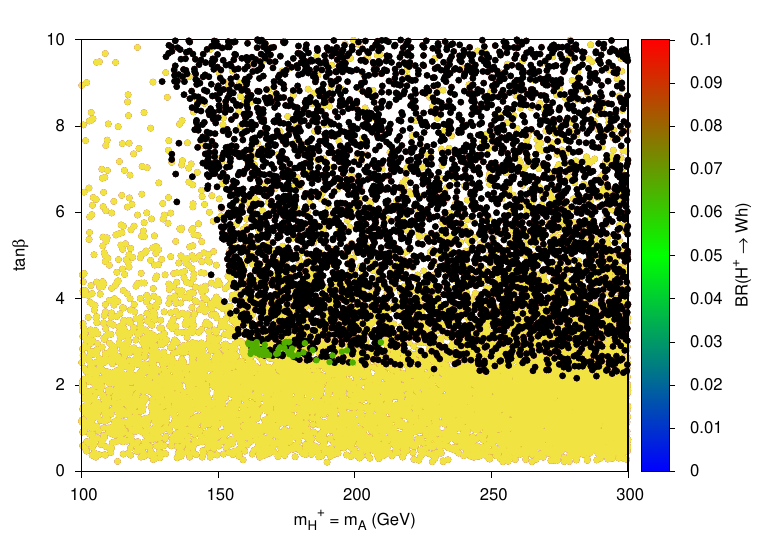}
\label{f1exptheoWh.pdf}
}\hfill
{\includegraphics[width=7.5cm]{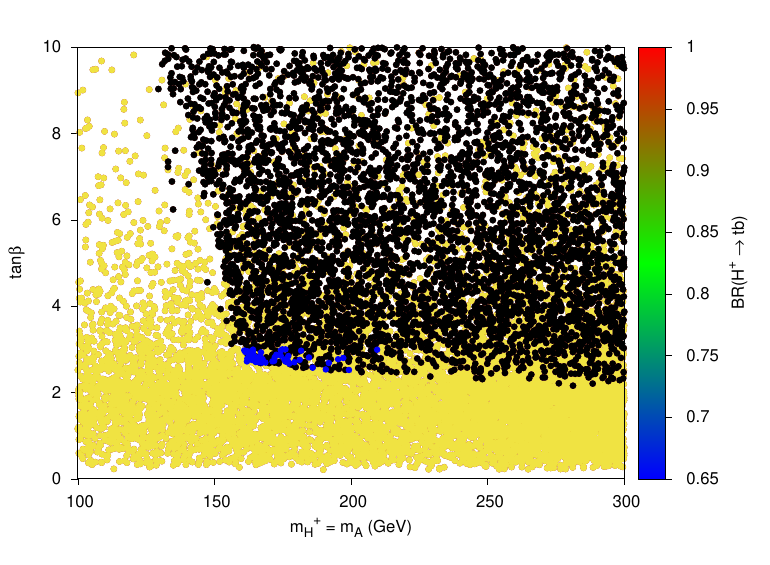}
\label{f1exptheotb.pdf}
}\hfill
\caption{In Type-I 2HDM, $m_{H^{+}}$ against $tan\beta$ BR($H^{+}\rightarrow W^{+}h$) (left) and BR($H^{+}\rightarrow t\bar{b}$) (right) is scanned in the presence of theoretical constraints as well as experimental constraints with $m_{h}$=125 GeV, $m_{H}$=300 GeV, $m_{H^{+}}$=$m_{A}$, $sin(\beta-\alpha)$=0.85, we set ${m^{2}}_{12}$=$\frac{[{m^2}_{H^+}][tan{\beta}]}{[1+{tan{\beta}}^2]}$. At a 95\% confidence-level (CL), yellow colour zones are omitted from LHC Higgs data, and black/grey zones are omitted from theoretical restrictions (constraints).}
\label{tanBvsmass3D}
\end{figure}

Figure \ref{BRWhandtbConstraints} (left) is the selection of points in ($tan\beta$,BR($H^{+}\rightarrow W^{+}h)$  parameter space when both theoretical and experimental constraints are applied at different charged Higgs masses while the Figure \ref{BRWhandtbConstraints} (right) is the  BR($H^{+}\rightarrow tb$)  parameter space. Effects of varying the parameters fixed here, will be shown and discussed later.

Figure \ref{tanBvsmass3D} shows that for \begin{math} tan{\beta} \in [2,3] \end{math} and \begin{math} m_{H^+} \in [150,210] \end{math}, the BR\begin{math}(H^{+} \rightarrow W^{+}h) \end{math} shows up in the region from [0.06,0.07] as shown in the vertical palette. It is also observed that for \begin{math} tan{\beta} \in [2,3] \end{math} and \begin{math} m_{H^+} \in [150,210] \end{math}, the BR\begin{math}(H^{+} \rightarrow tb) \end{math} shows up in the region from [0.65,0.7] as shown in the vertical palette. Under a same range of \begin{math} m_{H^{+}} \end{math} and $tan\beta$, BR\begin{math}(H^{+}\rightarrow tb) \end{math} becomes boosted towards the point that it overrides all the other decay modes. When going over light-charged Higgs to heavy-charged Higgs, we see there are most observable states that exist in the defined region.

\begin{figure}[!hbt]

{\includegraphics[width=7.5cm]{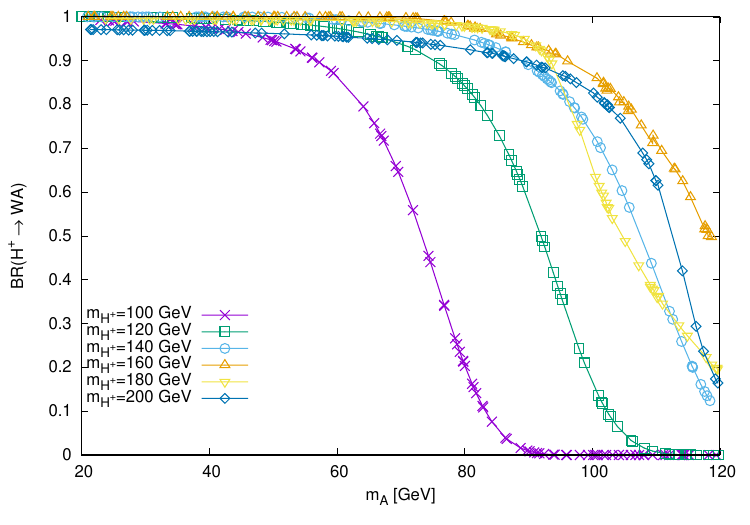}
\label{BP2100200thWA.pdf}
}\hfill
{\includegraphics[width=7.5cm]{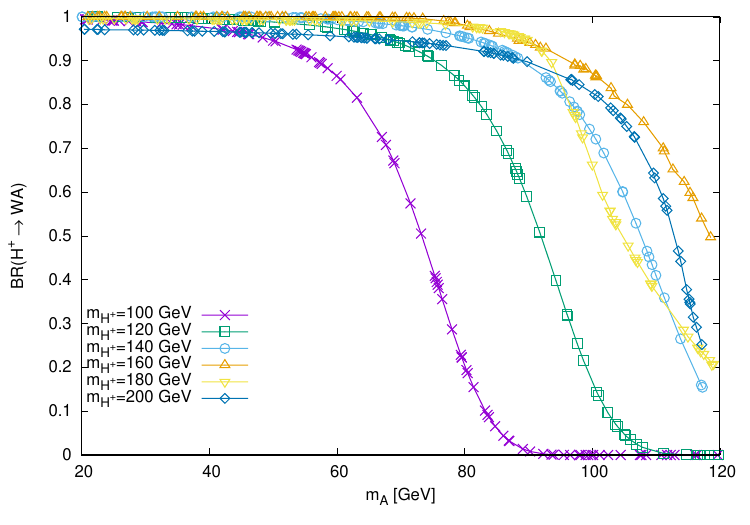}
\label{BP2100200exWA.pdf}
}\hfill
\caption{In Type-I 2HDM, $m_A$ vs BR($H^{+}\rightarrow W^{+}A$) is scanned in the face of theoretical (left) and experimental (right) restrictions (constraints) with $m_{h}$=125 GeV, $m_{H}$=300 GeV, $m_{H^{+}}$=[100-200] GeV, $tan\beta$=5, $sin(\beta-\alpha)=0.85$ and ${m^{2}}_{12}$=16000 Ge$V^{-2}$.}
\label{BRvsmAWh}
\end{figure}

\begin{figure}[!hbt]

{\includegraphics[width=7.5cm]{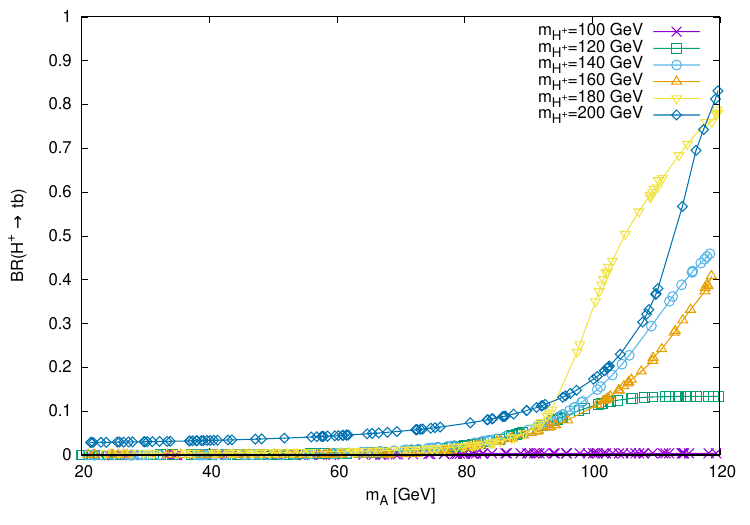}
\label{BP2100200thtb.pdf}
}\hfill
{\includegraphics[width=7.5cm]{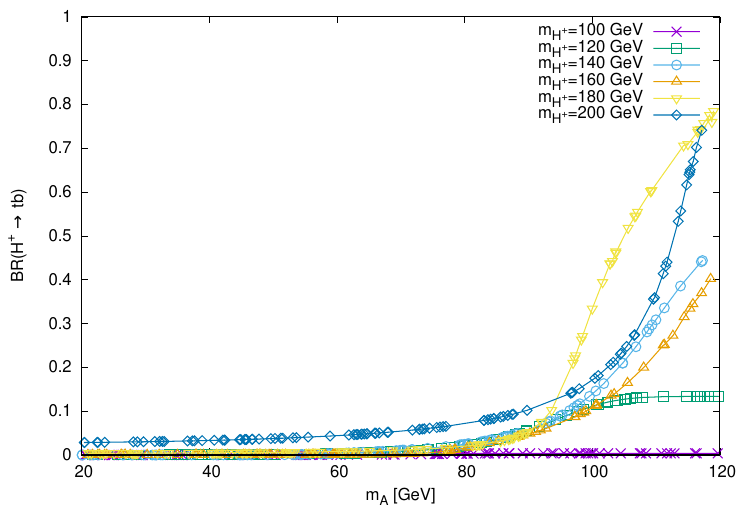}
\label{BP2100200extb.pdf}
}\hfill
\caption{In Type-I 2HDM, $m_A$ vs BR($H^{+}\rightarrow$ tb) is scanned in the face of theoretical (left) and experimental (right) restrictions (constraints) with $m_{h}$=125 GeV, $m_{H}$=300 GeV, $m_{H^{+}}$=[100-200] GeV, $tan\beta$=5, $sin(\beta-\alpha)=0.85$ and ${m^{2}}_{12}$=16000 Ge$V^{-2}$.}
\label{BRvsmAtb}
\end{figure}

From Figure \ref{BRvsmAWh} and Figure \ref{BRvsmAtb}, it is observed that the experimental bounds do not pose any further constraints  for $H^{+}\rightarrow W^{+}A$ and $H^{+}\rightarrow tb$.


\begin{figure}[!hbt]

{\includegraphics[width=7.5cm]{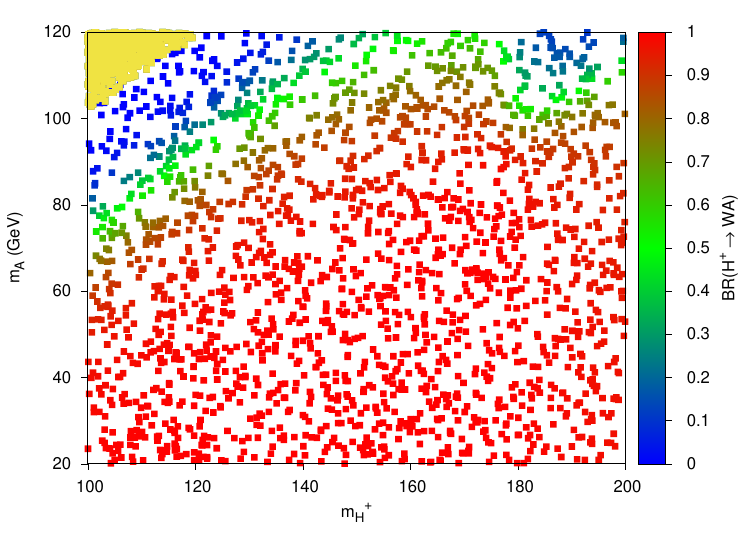}
\label{fig3exptheoWA.pdf}
}\hfill
{\includegraphics[width=7.5cm]{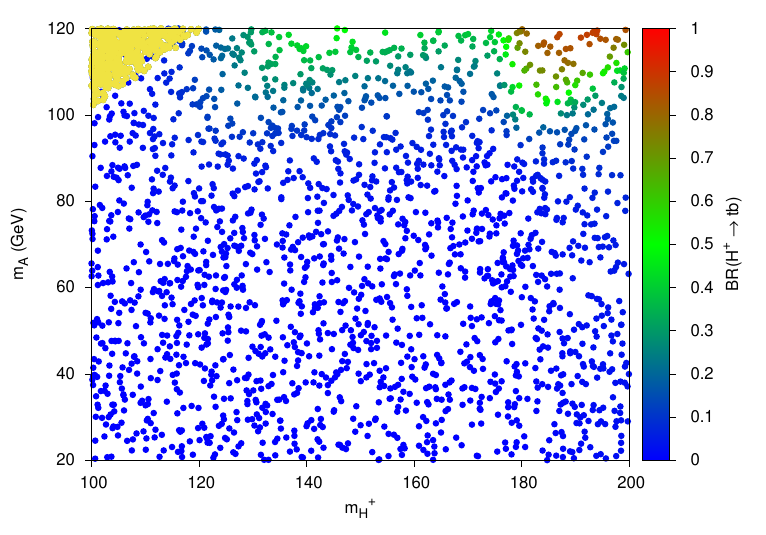}
\label{fig3exptheotb.pdf}
}\hfill
\caption{In Type-I 2HDM, $m_{H^{+}}$ vs $m_{A}$ BR($H^{+}\rightarrow WA$) (left) and BR($H^{+}\rightarrow$ tb) (right) is scanned in face of theoretical as well as experimental restrictions (constraints) with $m_{h}$=125 GeV, $m_{H}$=300 GeV, $tan\beta$=5, $sin(\beta-\alpha)=1$ and ${m^{2}}_{12}$=16000 Ge$V^{-2}$. At a 95\% confidence level (CL), the yellow zone is omitted from the LHC Higgs data.}
\label{mAvsmHplus3D}
\end{figure}

Figure \ref{mAvsmHplus3D}, shows that except some short regions, the BR\begin{math}(H^{+}\rightarrow WA) \end{math} is dominant and becomes 100\% in the mass range \begin{math} m_{H^+} = [100,200] \end{math}. The tb-decay mode is the least dominant because of the kinematic constraints and it only becomes prominent  for $m_{H^+}$ above 180 GeV. It is clear that BR\begin{math}(H^{+}\rightarrow WA) \end{math} could be the leading decay channel, i.e, for \begin{math} m_{A} \leq 100 \end{math}\hspace{1mm} GeV and any mass of $m_{H^+}$.

\begin{figure}[!hbt]

{\includegraphics[width=7.5cm]{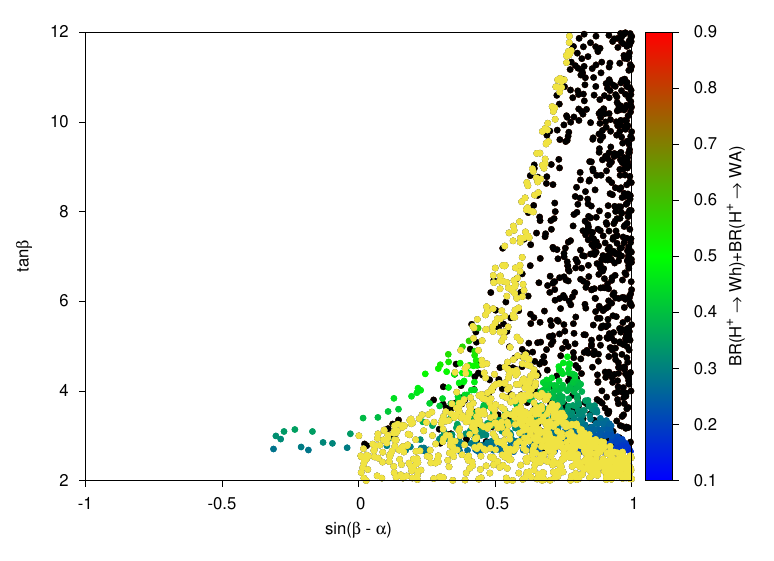}
\label{sinba.pdf}
}\hfill
{\includegraphics[width=7.5cm]{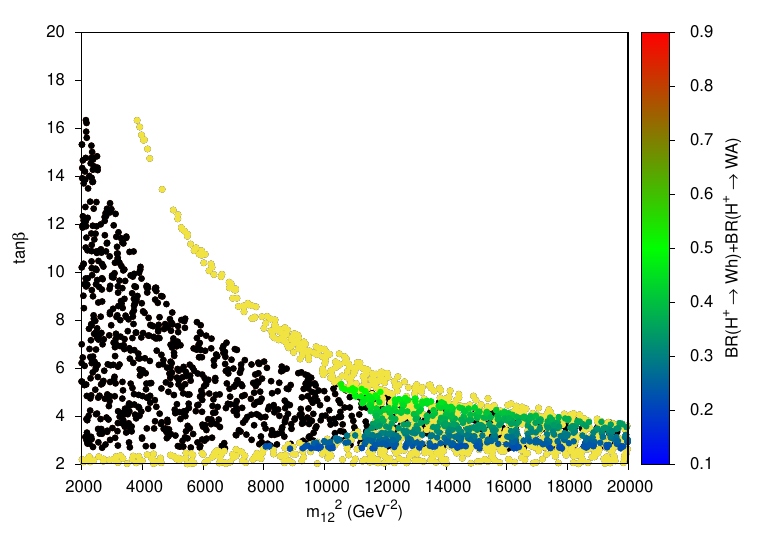}
\label{m212.pdf}
}\hfill
\caption{In THDM Type-I, BR($H^{+}\rightarrow W^{+}h$)+BR($H^{+}\rightarrow W^{+}A$) is mapped over $sin(\beta-\alpha$) vs $tan\beta$ with ${m^{2}}_{12}$=5000 $GeV^{-2}$ on the left while ${m^{2}}_{12}$ vs $tan\beta$ with $sin(\beta-\alpha)$=0.65 is shown on the right. Other parameters are $m_{h}$=$m_{A}$=125 GeV, $m_{H}$=300 GeV and $m_{H^{+}}$=170 GeV. At a 95\% confidence-level (CL), yellow colour zones are omitted from LHC Higgs data, and black/grey zones are omitted from theoretical restrictions (constraints).}
\label{tanBvsSinBmABR}
\end{figure}

Figure \ref{tanBvsSinBmABR} shows a scan over arbitrarily chosen and fixed $m_{H^{+}}$ between 160 GeV and 180 GeV i.e. 170 GeV. It shows the size of BR($H^{\pm}\rightarrow W^{\pm{*}}h + W^{\pm{*}}A$) over the $sin(\beta-\alpha)$ vs $tan\beta$ (left) and ${m^{2}}_{12}$ vs $tan\beta$ (right). The right panel shows the effect of the soft $Z_2$ breaking term $m_{12}$. It is clear that for some special choices of $tan\beta$ and ${m^{2}}_{12}$, the BR($H^{\pm}\rightarrow W^{\pm{*}}h + W^{\pm{*}}A$) could reach above 50\% for $sin(\beta - \alpha)$ between 0.6 and 0.7. The right figure is drawn by fixing $sin(\beta - \alpha)$ at 0.65, and variation is shown as a function of $m_{12}^2$. The favorable green zones are located in regions with $m_{12}^2 > 12,000$ GeV.


\begin{figure}[!hbt]

{\includegraphics[width=7.5cm]{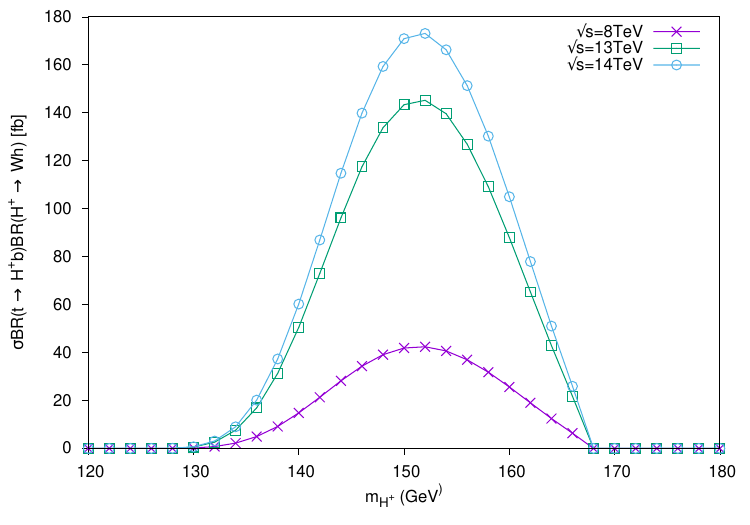}
\label{CSWh1.pdf}
}\hfill
{\includegraphics[width=7.5cm]{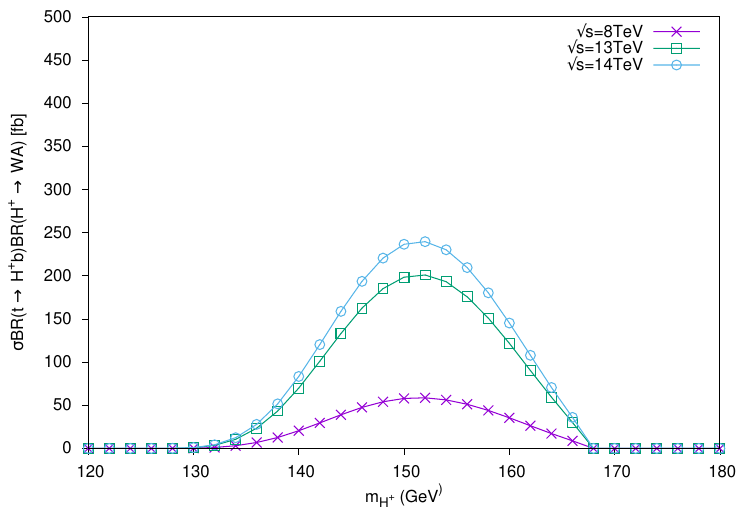}
\label{CSWA1.pdf}
}\hfill
\caption{In Type-I 2HDM, the rates for $\sigma(pp\rightarrow t\bar{t})$ $\times$ $BR(t\rightarrow H^{+}b) \times BR(H^{+}\rightarrow W\phi)$ with $\phi=h$ (left) and $\phi=A$ (right) are scanned as a function of $m_{H^{+}}$ at $\sqrt{s}$=8 TeV, $\sqrt{s}$=13 TeV and $\sqrt{s}$=14 TeV with $m_{h}$=125 GeV, $m_{H}$=300 GeV, $m_{A}$=125 GeV, $tan\beta$=5, $sin(\beta-\alpha)$=0.85 and ${m^{2}}_{12}$=16000 Ge$V^{-2}$.}
\label{sigmaBRvsmHp}
\end{figure}

Figure \ref{sigmaBRvsmHp}, shows $\sigma \times$ BR($t\rightarrow H^{+}b$) $\times$ BR($H^{+}\rightarrow W\phi$) with $\phi=h$ (left) and $\phi=A$ (right) over light charged Higgs mass range 120-180 GeV in collisions of proton and proton at three distinct centre-of-mass energies i.e. $\sqrt{s}$ = 8 TeV, $\sqrt{s}$ = 13 TeV and $\sqrt{s}$ = 14 TeV with three different distinguishable colors.  

\begin{figure}[!hbt]

{\includegraphics[width=7.5cm]{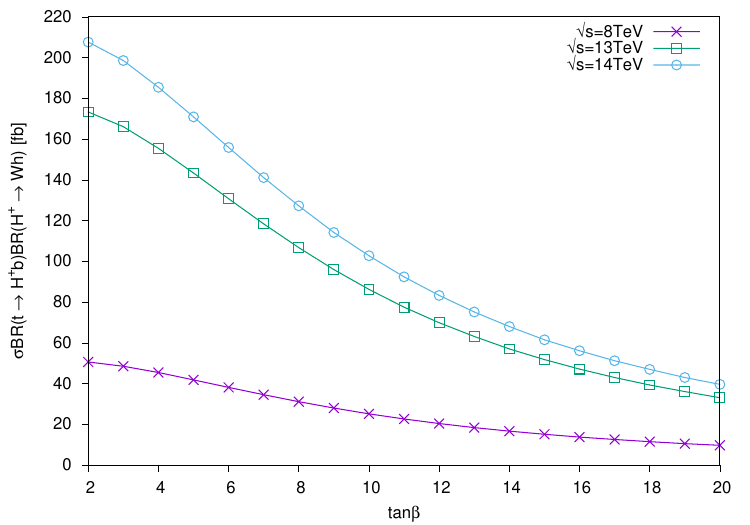}
\label{tbvsCSWh.pdf}
}\hfill
{\includegraphics[width=7.5cm]{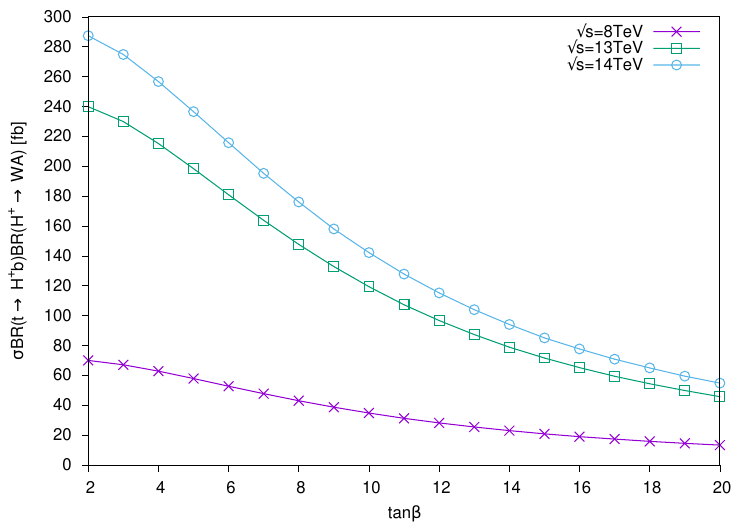}
\label{tbvsCSWA.pdf}
}\hfill
\caption{In Type-I 2HDM, the rates for $\sigma(pp\rightarrow t\bar{t})$ $\times$ $BR(t\rightarrow H^{+}b) \times BR(H^{+}\rightarrow W\phi)$ with $\phi=h$ (left) and $\phi=A$ (right) are scanned as a function of $tan\beta$ at $\sqrt{s}$=8 TeV, $\sqrt{s}$=13 TeV and $\sqrt{s}$=14 TeV with $m_{h}$=125 GeV, $m_{H}$=300 GeV, $m_{H^{+}}$=170 GeV, $m_{A}$=125 GeV, $sin(\beta-\alpha)$=0.85 and ${m^{2}}_{12}$=16000 Ge$V^{-2}$.}
\label{sigmaBRtanB}
\end{figure}

Figure \ref{sigmaBRtanB} shows the variation of $\sigma \times$ BR($t\rightarrow H^{+}b$) $\times$ BR($H^{+}\rightarrow W\phi$) as a function of $tan \beta$, ranging between 2-20, while keeping the charged higgs mass $m_H^{\pm}$ fixed at 170 GeV. Three distinct curves represent different centre of mass energies, as described earlier.


\section{Conclusion} 
In this study, Type-I 2HDM is used as the theoretical foundation, the scenario selected is similar to the standard model, with the lighter scalar Higgs (h) acting as the SM Higgs boson, in search for a potential discovery channel for the light-$H^{+}$ via bosonic decay channels. Bosonic decays of $H^{+}$, are investigated explicitly, as this mode has not been explored in-depth till now. All $m_{H^{+}}$ values are taken such that they are allowed kinematically.  As it is observed when $m_A$=$m_{H^{+}}$ and $sin(\beta-\alpha)$ fixed at 0.85 then for $m_{H^{+}}$=400 GeV at 9$\leq tan\beta \leq$ 10, BR($H^{+}\rightarrow W^{+}h$) approximately 100\% while at $m_{H^{+}}$=160 GeV maximum of 20\%-30\% of branching ratio observed. 
  
For higher masses of charged Higgs boson, $H^{+}\rightarrow W^{+}A$ decay channel becomes less dominant but it is observed at 160 GeV and 180 GeV of charged Higgs mass becomes most dominant when $m_{A}$ varies between 20 and 120 GeV and $sin(\beta-\alpha)$ varies in the range [-1,1]. When $m_{H^{+}}$ fixed arbitrarily at 170 GeV then it can clearly be observed that BR($H^{+}\rightarrow W^{+}h$)+BR($H^{+}\rightarrow W^{+}h$) becomes up to 50\% observable $m_A$=$m_{h}$=125 GeV and fixing $sin(\beta-\alpha)$ at 0.65. The highest value of $\sigma$ is observed at $m_{H^{+}}$=150 GeV at 8 TeV, 13 TeV and 14 TeV of $\sqrt{s}$ in $pp$-collisions. So by fixing $m_{H^{+}}$=150 GeV, the highest value of $\sigma$ is observed at $tan\beta$=2 and lowest at $tan\beta$=20 for all three different values of $\sqrt{s}$. It is also observed that with an increase of $tan\beta$, the value of cross-section decreases for $H^{+}\rightarrow W\phi$ (where $\phi$=h or A). Several scans are performed for observing bosonic decay modes of $H^{\pm}$ from $[200-1000]~GeV$ mass range and then confining the range 100-200 GeV and in the next step fixing it at 170 GeV for both decay modes ($W^{+}h$, $W^{+}A$) by applying theoretical bounds and experimental bounds from most recent Higgs searches. It can be seen that the possibility exists for light $H^{+}$ to decay via $H^{+}\rightarrow W^{+}h$ channel or $H^{+}\rightarrow W^{+}A$ channel (bosonic decay modes). These scans are for observing and calculating branching ratios and cross-sections for different bosonic decay modes and making a comparison with top-bottom decay modes. This study shows that light charged Higgs boson is possibly observable via bosonic decay channels and it leads the experimentalists to an alternative possible discovery channel for light charged Higgs boson. This study provides a specifically good way to examine beyond standard model(BSM) Higgs bosons as well as validate the sustainability and feasibility of the 2HDM in the selected parameter space.\\
\textbf{Availability of data and materials}\\
Data sharing not applicable to this article as no datasets were generated or analysed during the current study.\\

 \bibliographystyle{ieeetr}
 
\end{document}